\documentclass[acmsmall,screen,nonacm]{acmart}

\setcopyright{none}
\usepackage{amsmath}
\usepackage{tabularx}
\usepackage{multirow}
\usepackage{listings}
\usepackage{xcolor}
\usepackage{tikz}
\usetikzlibrary{arrows.meta,backgrounds,calc,fit,positioning}
\newcommand{\compacttable}[1][0.95]{%
  \footnotesize
  \renewcommand{\arraystretch}{#1}%
  \setlength{\tabcolsep}{3pt}%
}
\lstdefinestyle{paperpatch}{
  basicstyle=\ttfamily\footnotesize,
  frame=single,
  framerule=0.4pt,
  framesep=3pt,
  xleftmargin=4pt,
  xrightmargin=4pt,
  columns=fullflexible,
  keepspaces=true,
  showstringspaces=false,
  upquote=true,
  tabsize=2,
  breaklines=true,
  breakatwhitespace=true,
  breakautoindent=false,
  breakindent=1em,
  postbreak=\mbox{\(\hookrightarrow\)\space},
  aboveskip=3pt,
  belowskip=3pt
}
\newcommand{\patchcontent}[3][\footnotesize]{%
  \footnotesize\textbf{#2}\par
  \lstset{style=paperpatch,basicstyle=\ttfamily#1,frame=none,xleftmargin=0pt,xrightmargin=0pt}%
  #3
}
\newcommand{\patchbox}[3][\footnotesize]{%
  \par\smallskip\noindent
  \begingroup\setlength{\fboxsep}{1pt}%
  \fbox{\begin{minipage}{\dimexpr\linewidth-2\fboxsep-2\fboxrule\relax}
    \patchcontent[#1]{#2}{#3}
  \end{minipage}}
  \endgroup
  \par\smallskip
}
\newsavebox{\patchleft}
\newsavebox{\patchright}
\newlength{\patchpairheight}
\newcommand{\pairedpatchboxes}[5][\scriptsize]{%
  \par\smallskip\noindent
  \begingroup\setlength{\fboxsep}{1pt}%
  \sbox{\patchleft}{\begin{minipage}{\dimexpr0.49\linewidth-2\fboxsep-2\fboxrule\relax}
    \patchcontent[#1]{#2}{#3}
  \end{minipage}}%
  \sbox{\patchright}{\begin{minipage}{\dimexpr0.49\linewidth-2\fboxsep-2\fboxrule\relax}
    \patchcontent[#1]{#4}{#5}
  \end{minipage}}%
  \setlength{\patchpairheight}{\dimexpr\ht\patchleft+\dp\patchleft\relax}%
  \ifdim\dimexpr\ht\patchright+\dp\patchright\relax>\patchpairheight
    \setlength{\patchpairheight}{\dimexpr\ht\patchright+\dp\patchright\relax}%
  \fi
  \fbox{\vbox to \patchpairheight{\hbox{\usebox{\patchleft}}\vfil}}\hfill
  \fbox{\vbox to \patchpairheight{\hbox{\usebox{\patchright}}\vfil}}%
  \endgroup
  \par\smallskip
}
\newcommand{\tracebox}[3][\footnotesize]{\patchbox[#1]{#2}{#3}}
\newcommand{\reviewcomment}[2]{%
  \par\smallskip\noindent
  \begingroup\setlength{\fboxsep}{1pt}%
  \fbox{\begin{minipage}{\dimexpr\linewidth-2\fboxsep-2\fboxrule\relax}
    \footnotesize\textbf{Example review comment}\par\smallskip
    \textbf{#1}\par\smallskip
    #2
  \end{minipage}}
  \endgroup
  \par\smallskip
}
\definecolor{answergray}{gray}{0.92}
\newcommand{\answerbox}[1]{%
  \par\smallskip\noindent
  \begingroup\setlength{\fboxsep}{6pt}%
  \colorbox{answergray}{%
    \begin{minipage}{\dimexpr\linewidth-2\fboxsep\relax}
      \small\textbf{Answer:}\par\smallskip
      #1
    \end{minipage}}%
  \endgroup
  \par\smallskip
}
\newcommand{\ImplementationCount}{339}
\newcommand{\SyntheticLabelCount}{882}

\newcommand{\InfoMapCount}{678}
\newcommand{\InfoQuestionCount}{4,928}
\newcommand{\InfoQuestionMedian}{8}
\newcommand{\InfoBeyondRequestCount}{4,866}
\newcommand{\InfoBeyondRequestPercent}{98.74}
\newcommand{\InfoToolBearingCount}{4,858}
\newcommand{\InfoToolBearingPercent}{98.58}

\title[Trajectory-Informed Code Review]{Diffs Are Cheap, Show Me the Trajectory: An Empirical Study of Trajectory-Informed Code Review}
\input{authors}

\begin{document}

% !TEX root = ../../../main.tex
\begin{abstract}
Code review typically starts from the final diff.
However, the diff reveals little about how the code was developed.
Coding-agent trajectories preserve
intermediate decisions, failed attempts, and validation evidence that may
assist code review. We systematically investigate whether and how these
trajectories can help independent review agents identify problems in the
resulting code, beyond inspecting the final diff and repository.
We develop a procedure that generates review comments by comparing the
initial and merged versions of pull requests, and assess it against
withheld human review comments. Using held-out reference implementations
as comparison targets for agent-produced code, we construct a dataset
linking \ImplementationCount{} implementations of 113 DeepSWE tasks with
their trajectories and \SyntheticLabelCount{} synthetic reference review
comments. Across three coding-agent configurations and two review
harnesses, we compare diff-only review, full-trajectory access, and four
alternative ways of presenting and using trajectories.
Full-trajectory access raises mean reference recall from 23.03\% to
25.87\%, but improves recall in only four of six configurations.
All four alternatives improve mean reference recall over diff-only by
4.05--7.32 percentage points. The gains also extend to review
comments that explain task-verifier failures. Higher recall can require
more tokens and longer review time, while simpler trajectory views
remain competitive. These results support development trajectories as an
additional source of code-review context and show that their benefits and
resource costs depend on how they are presented and used.
\end{abstract}
\maketitle
\raggedbottom

% !TEX root = ../../../main.tex
\section{Introduction}
\label{sec:introduction}

Code review requires understanding a change's intended behavior and its
effects on the surrounding code~\cite{bacchelli2013review}.
Automated code review systems supplement patches with retrieved examples,
pull-request metadata, and repository
context~\cite{li2022auger,zhang2025laura,sun2026gocodereview}.
These sources help reviewers understand the resulting implementation.
However, the final patch does not preserve the full process that produced
it: alternatives may have been discarded, assumptions revised, and tests
added in response to earlier failures. Such information can help explain
why a change took its current form and where further checks may be needed.

Coding agents make these aspects of development observable through their
development trajectories. Their workflows span repository exploration,
implementation, debugging, and testing~\cite{yang2024sweagent,wang2024openhands}.
These trajectories contain messages, tool calls, revisions, and validation attempts.
A discarded approach may expose a compatibility constraint, while a
failed test and its subsequent repair may suggest related scenarios that
remain unchecked. These trajectories therefore offer potential review leads
beyond what is visible in the final patch.

Access to a trajectory, however, does not automatically make its contents
useful for review. Relevant observations are interleaved with lengthy
command output, superseded code, and assumptions that may be wrong.
A reviewer must identify which parts matter and check their implications
against the final implementation. The question is therefore not only
whether development trajectories improve review, but also how reviewers
should select, organize, and consult this process evidence.

We study \emph{trajectory-informed code review}, in which an independent
review agent examines a coding agent's final diff and repository, with
access to the trajectory from the same development run.
Our focus is whether this additional information source helps
the reviewer identify problems in the resulting code, and under what
conditions it is useful. We first compare full-trajectory access with
\emph{diff-only review}, where the reviewer examines the patch and
repository without the development trajectory. We then investigate four ways
of presenting and using trajectory information: Messages, Self-selected
signals, Evidence-linked map, and Code-first adversarial scrutiny.
These configurations vary which information is retained, how it is
organized, and when it is consulted during review. Together, they allow
us to examine the value of development-process context across different
ways of incorporating it into review, rather than through a single
method.

Evaluating these approaches requires code changes and trajectories paired
with reference review comments. Our initial inspection of public projects
with available coding-agent trajectories found sparse human review feedback.
We address this data gap with a procedure that compares the initial and
merged versions of a pull request. Guided by the task requirements, it
uses the differences between these versions to generate review comments
about the initial version. We assess the procedure on human-reviewed
pull requests, withholding the human review comments during generation.
This evaluation shows that the procedure can recover concerns raised
by human reviewers.

We then run coding agents under three model--harness configurations on 113 DeepSWE
tasks~\cite{datacurve2026deepswe} to collect implementations and development
trajectories. Comparing these implementations with the tasks' held-out
reference implementations yields synthetic review comments that serve as
evaluation references. The resulting dataset contains
\ImplementationCount{} implementations, their trajectories, and
\SyntheticLabelCount{} reference review comments.
We review the implementations through Codex CLI and GitHub Copilot CLI
in \texttt{/review} mode, both using GPT-5.6 Sol, yielding six
coding--review configurations.
Paired comparisons use the same implementations and reference review
comments. We measure \emph{reference recall}, the proportion of reference
review comments recovered, together with token use, review time, and
estimated API cost. Our evaluation addresses three research questions:
\begin{description}
  \item[RQ1:] Does access to the full development trajectory improve code review over diff-only review?
  \item[RQ2:] Which representations and review procedures best support trajectory-informed code review?
  \item[RQ3:] What are the impacts of trajectory-informed review on cost and critical failure-linked coverage?
\end{description}

Across the six configurations, all four trajectory-informed methods
improve mean reference recall over diff-only by 4.05--7.32 percentage
points. Full-trajectory access alone raises mean reference recall from
23.03\% to 25.87\%, but improves recall in only four configurations.
All four methods also improve mean recall on \emph{critical failure-linked}
review comments, which explain DeepSWE verifier failures.
The methods differ in both coverage and resource use: the evidence-linked
map achieves the highest mean recall at 30.35\%, while Messages reaches
29.03\% with fewer tokens in all six configurations.
These results support using development trajectories as an additional
source of review context, while showing that their benefits and costs
depend on how that context is presented and used.

This paper makes three contributions:
\begin{itemize}
  \item \textbf{A study of development trajectories as code-review context.}
  We systematically investigate supplying coding agents' development
  trajectories to independent review agents, assessing their value beyond
  the final patch and repository context for reviewing the resulting code.
  \item \textbf{A trajectory-linked code-review dataset and construction procedure.}
  We link agent-produced implementations with development trajectories,
  synthetic reference review comments, and review outputs, and assess the
  comment-generation procedure against human review comments.
  \item \textbf{An empirical comparison of trajectory-use strategies.}
  We compare full-trajectory access and four alternative representations and
  review procedures across three coding configurations and two reviewer
  harnesses, examining reference coverage, critical failure-linked coverage,
  and resource use.
\end{itemize}

% !TEX root = ../../../main.tex
\section{Background and Motivation}
\label{sec:motivation}

We first define the contents of a development trajectory, then use a coding
agent's execution of a code-generation task from DeepSWE to illustrate how
the recorded development trajectory can guide code review.

\subsection{Development Trajectories}
\label{sec:motivation-trajectory}

A \emph{development trajectory} records how a coding agent carries out a
user request through successive interactions with its environment. Its
typical structure consists of an initial user request, repeated agent--tool
interactions, and a completion message:
\[
\tau = \langle u_0,\,
(m_1, c_1, o_1),\, \ldots,\,
(m_k, c_k, o_k),\, m_{\mathrm{final}} \rangle.
\]
Here, $u_0$ denotes the user request that specifies the coding task. Each
of the $k$ interaction rounds contains agent messages $m_i$, tool calls
$c_i$ with their arguments, and the corresponding tool observations $o_i$.
The final agent message $m_{\mathrm{final}}$ summarizes the implementation
and its validation.

Unlike the final patch, which describes the resulting code changes, the
trajectory can preserve intermediate attempts, failures, revisions, and
validation activities. The order of interaction rounds and the association
between calls and observations reveal how the implementation evolved.
For example, a failed test followed by an edit and a successful rerun
records a failure--revision--validation sequence. Such sequences help
reviewers trace implementation decisions, changing assumptions, and
unresolved questions. Table~\ref{tab:trajectory-contents} summarizes the
trajectory contents and their potential uses in review.

\begin{table}[t]
  \caption{Development trajectory contents and their potential review uses.}
  \label{tab:trajectory-contents}
  \compacttable
  \begin{tabularx}{\linewidth}{@{}p{0.26\linewidth}XX@{}}
    \toprule
    Content & Typical records & Potential review use \\
    \midrule
    User request ($u_0$) & Task description and constraints &
      Establish expected behavior. \\
    Agent messages ($m_i$, $m_{\mathrm{final}}$) & Plans, design rationale, assumptions, and completion claims &
      Identify implementation judgments to verify. \\
    Tool calls ($c_i$) & Searches, file reads, edits, and test commands with their arguments &
      Identify the actions taken and their scope. \\
    Tool observations ($o_i$) & Retrieved code, command errors, and test output &
      Examine execution evidence, failures, and validation limits. \\
    \bottomrule
  \end{tabularx}
\end{table}

Agent messages and tool observations play complementary roles in review:
messages expose the agent's intentions, assumptions, and claims, while
observations show the results of its actions. Comparing the two can reveal
assumptions that were not tested or claims that exceed the recorded checks,
providing concrete leads for verification against the final patch and repository.

% Figure~\ref{fig:trajectory-composition} illustrates how these contents differ
% in volume in a recorded trajectory. The complete serialized record for the
% GPT-5.5 run on \texttt{abs-module-cache-flags} is 304.43 KB. Tool observations
% occupy \TraceToolResultShare{}\% of its bytes, whereas agent messages occupy
% only \TraceAgentMessageShare{}\%. The record also contains repeated arguments,
% patch arguments, and serialization overhead. The second panel groups paired
% shell-command calls and observations by command family, using a different
% denominator. These shares describe this run, not a cohort average or
% token-cost estimate. Byte volume does not identify which events matter to a
% reviewer: a lengthy observation may contain a crucial failure, whereas a
% short message may expose an important assumption.

% \begin{figure}[t]
%   \centering
%   \includegraphics[width=\linewidth]{variants/fixed-reference/figures/trajectory-composition.pdf}
%   \caption{Byte composition of the GPT-5.5 run on \texttt{abs-module-cache-flags}. The
%   panels use different denominators: the full serialized trajectory record and
%   paired shell-command invocation/result objects, respectively.}
%   \Description{The left panel shows the serialized field shares, including
%   48.91 percent tool results and 2.64 percent agent messages. The right
%   panel groups shell invocation and result bytes by command family.
%   Both panels describe the GPT-5.5 run on abs-module-cache-flags,
%   not a cohort average.}
%   \label{fig:trajectory-composition}
% \end{figure}

\subsection{Motivating Example}
\label{sec:motivation-cattrs}

We illustrate how a development trajectory can guide review using a
DeepSWE task for cattrs\footnote{\url{https://deepswe.datacurve.ai/data/v1.1/tasks/cattrs-partial-structuring-recovery}},
a Python library that converts dictionaries and other unstructured data
into typed objects. For example, it can turn a dictionary containing
the string \texttt{"1"} into an \texttt{Inner} object whose integer field
\texttt{x} is \texttt{1}. Developers can customize this conversion by
registering a \emph{hook}: a function that cattrs should call to construct
objects of a particular type.

The user request $u_0$ asks the coding agent to add \emph{partial structuring}:
if some fields cannot be converted, keep the valid fields and report the
conversion errors. Codex CLI, backed by GPT-5.5, implements this task.
In the resulting trajectory $\tau$, an agent message $m_i$ explains that
the agent will add a separate conversion path rather than modify the
existing one. It assumes that the new path will still call registered
hooks, preserving developers' custom conversion behavior:

\par\smallskip\noindent
\begingroup\setlength{\fboxsep}{1pt}%
\fbox{\begin{minipage}{\dimexpr\linewidth-2\fboxsep-2\fboxrule\relax}\small
\textbf{Agent message: autonomous design decision (step 8)}\par
The generated fast paths are optimized for all-or-nothing construction, so I’m going to add the partial algorithm alongside \texttt{structure} rather than trying to alter every generated hook. It can still delegate individual field conversion to the existing registered hooks, which preserves custom hook behavior and collection atomicity.
\end{minipage}}
\endgroup\par\smallskip

When the development trajectory $\tau$ is provided to a review agent
alongside the code changes, the agent can use the recorded implementation
assumptions to identify what to check. Here, the message $m_i$ suggests a
concrete question: does the new partial-structuring path still call
registered hooks? In the recorded review, the review agent checks this
behavior by registering a hook for \texttt{Inner} that ignores the input
and always returns
\texttt{Inner(999)}. The value \texttt{999} is a marker: even when the
input contains \texttt{"1"}, calling the hook should produce an object
whose field is \texttt{999}. The code below shows a simplified version
of this check:

\patchbox{Code example: preserving a registered conversion hook}{\begin{lstlisting}
from attrs import define
from cattrs import Converter
@define
class Inner: x: int
c = Converter()
c.register_structure_hook(Inner, lambda o, _: Inner(999))
c.structure({"x": "1"}, Inner)  # Inner(999)
# Expected: Inner(999); observed: Inner(1)
c.partial_structure({"x": "1"}, Inner).value
\end{lstlisting}
}

The existing \texttt{structure} method returns \texttt{Inner(999)}, showing
that it calls the hook. The object returned by \texttt{partial\_structure},
accessed through \texttt{.value}, should be \texttt{Inner(999)} as well.
Instead, the recorded review probe returns \texttt{Inner(1)}.
The new path converts the field from \texttt{"1"} to \texttt{1} directly,
bypassing the hook for \texttt{Inner}. Although this looks like a valid
conversion, it ignores the developer's customization and contradicts the
assumption expressed in $m_i$.

The trajectory $\tau$ provides a specific lead for review: the coding
agent claims that custom hooks will still work, and the review agent's
check reveals that they do not work on the new path. The message $m_i$
helps identify what to check; executing the check supplies the evidence
of the defect. This motivates our study of whether development trajectories
improve code review across tasks, how to represent and use them, and at
what additional cost.
% !TEX root = ../../../main.tex
\section{Methodology}
\label{sec:methodology}
\label{sec:dataset}

To evaluate whether development trajectories improve code review, we
need the code changes and development trajectory for each coding task,
along with reference review comments against which to evaluate review outputs.
We explored projects with agent sessions recorded through Entire.io,
as represented in SWE-chat~\cite{baumann2026swechat}, but our preliminary
inspection found sparse human review comments on the corresponding
changes. We therefore construct a dataset that links these three
components for each coding task.

We first establish how to generate review comments
(Section~\ref{sec:reference-construction}). We use the differences between
an initial implementation and its merged version to generate review comments
about the initial implementation, then compare them with withheld human
review comments. This checks how well the generated review comments correspond
to human review comments.

\begin{figure*}[t]
  \centering
  \includegraphics[width=\textwidth]{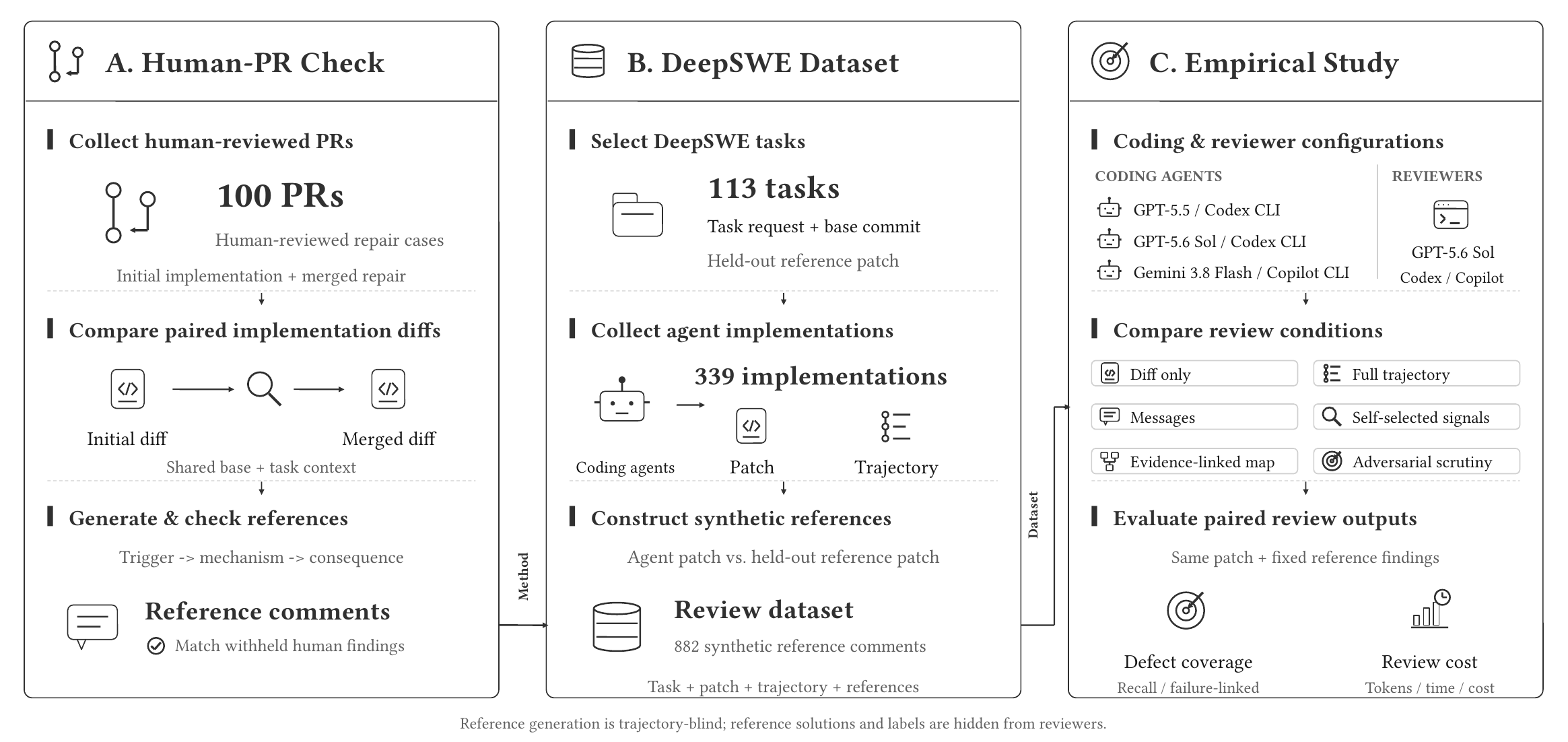}
  \caption{Dataset construction and trajectory-informed review workflow.}
  \Description{A monochrome workflow has three columns, each with three
  subsections. The human-PR column compares initial and merged diffs and
  checks generated reference comments against withheld human findings.
  The DeepSWE column collects agent patches and trajectories and constructs
  synthetic references. The study column shows coding and reviewer
  configurations, six review conditions, and paired coverage and cost
  evaluation. Arrows transfer the construction method and resulting dataset.}
  \label{fig:framework}
\end{figure*}

We then run coding agents on DeepSWE tasks to collect implementations
and development trajectories, and generate synthetic review comments
using the tasks' held-out reference implementations
(Section~\ref{sec:trajectory-dataset}). These generated comments serve as
\emph{reference review comments} for evaluating review outputs.
Finally, we describe the experimental setup and evaluation metrics used
to study trajectory-informed code review on this dataset
(Section~\ref{sec:study-design}).
Figure~\ref{fig:framework} summarizes this workflow.

\subsection{Reference Review Comment Construction}
\label{sec:reference-construction}

\subsubsection{Construction Procedure}
\label{sec:diff-comparison}

For a human-reviewed pull request, let $B$ denote the repository state
immediately before the initial implementation commit. Let $\Delta_I$
and $\Delta_M$ denote the changes from $B$ to the initial implementation
and the merged version, respectively. The two resulting program states
are:
\[
S_I = B \oplus \Delta_I,
\qquad
S_M = B \oplus \Delta_M,
\]
where $\oplus$ denotes applying a patch to the repository.

Our goal is to generate a set of review comments
\[
F = \{f_1, f_2, \ldots, f_k\},
\]
where each comment $f_i$ describes a concern about $S_I$ suggested by
the changes leading to $S_M$. The merged version provides evidence of
what changed; the model uses this evidence to explain why a reviewer
might request that change in the initial version.

We provide a language model with the recorded user request and the two
patches, $\Delta_I$ and $\Delta_M$, but not the human review comments.
The model compares the two versions and generates comments explaining
the relevant code, the circumstances in which a concern arises, its
possible consequences, and how the later change addresses it. The prompt
also asks the model to consider expected behavior, code that uses the
changed functionality, boundary cases, and potential regressions.

For example, Alamofire\footnote{\url{https://github.com/Alamofire/Alamofire/pull/3486}},
a Swift networking library, uses conditional compilation to control
its integration with Apple's Combine framework. In the initial
implementation, the compilation guard excludes 32-bit iOS but not
Linux or Windows. On these two platforms, the guard evaluates to
true, allowing compilation to reach an import of Combine even when
the framework is unavailable. The merged version extends the guard
to exclude Linux and Windows:

\pairedpatchboxes[\scriptsize]
  {Initial patch}{\begin{lstlisting}
-#if canImport(Combine)
+#if !(os(iOS) && (arch(i386) || arch(arm)))
 
 import Combine
\end{lstlisting}
}
  {Merged patch}{\begin{lstlisting}
-#if canImport(Combine)
+#if !((os(iOS) && (arch(i386) || arch(arm))) || os(Windows) || os(Linux))
 
 import Combine
\end{lstlisting}
}

From this change, our approach generates the following review comment,
which explains the potential build failure and the corresponding correction,
matching the platform-exclusion concern raised in the withheld human
review comment:

\reviewcomment{Explicitly exclude Linux and Windows from the Combine guards}{On these platforms, os(iOS) is false, so the current guard evaluates to true and reaches import Combine. This can break builds where Combine is unavailable. Apply the exclusions to both the source and the Combine tests.}

\subsubsection{Comparison with Human Review Comments}
\label{sec:human-corpus}

We check how well the generated review comments correspond to human
review comments on the same code changes. We collect public, non-fork,
non-archived GitHub repositories using
an MIT-license filter and a default minimum of 10,000 stars.
These established projects provide a practical source of human-reviewed
code changes. We retain merged, human-authored PRs with exactly two
linear commits: the first introduces the initial implementation,
and the second contains the version ultimately merged. Both versions
must be reconstructable from complete patches.
Each retained PR must concern a single requested change. We exclude automated,
documentation-only, and generated-file changes, as well as unrelated
maintenance, refactoring, or formatting.

We collect human review comments from root review-thread comments
written by reviewers other than the PR author, after the initial
implementation commit and before the second commit. Comment timestamps,
code locations, and the corresponding code changes are used to associate
the comments with the paired versions. The resulting corpus contains
100 PRs from 49 repositories and 126 human review comments.

We use GPT-5.5 at medium reasoning effort to generate review comments
from the user request and paired patches, without providing the human
comments as input. We then use semantic matching to compare the generated
and human review comments, checking whether they raise the same concern
rather than requiring identical wording.

We repeat review-comment generation and evaluation independently three
times on the same 100 PRs. Table~\ref{tab:reference-generation} reports
each round and the three-round mean.

As shown in Table~\ref{tab:reference-generation}, our approach recovers
102.33 of 126 human review comments on average, yielding
81.22\% mean micro recall. Within each round, we compute micro recall by pooling
matched and total human review comments across all
100 PRs before division, rather than averaging per-PR recall.
The approach generates 113.67 review comments on average and achieves
100.00\% mean micro precision and 89.62\% mean micro F1.
These results show that comparing initial and merged versions can
generate review comments that correspond to a substantial share of
the human review comments in this corpus.

% !TEX root = ../../../../main.tex
\begin{table}[t]
  \caption{Recovery of human review comments from initial and merged implementations over three independent rounds.}
  \label{tab:reference-generation}\compacttable
  \begin{tabular*}{\textwidth}{@{\extracolsep{\fill}}lrrrrrr@{}}
    \toprule
     & \shortstack[c]{Generated\\review\\comments} & \shortstack[c]{Human\\review\\comments} & \shortstack[c]{Matched human\\review\\comments} & \shortstack[c]{Micro\\precision\\(\%)} & \shortstack[c]{Micro\\recall\\(\%)} & \shortstack[c]{Micro F1\\(\%)} \\
    \midrule
    Round 0 & 118 & 126 & 106 & 100.00 & 84.13 & 91.38 \\
    Round 1 & 112 & 126 & 100 & 100.00 & 79.37 & 88.50 \\
    Round 2 & 111 & 126 & 101 & 100.00 & 80.16 & 88.99 \\
    \midrule
    Mean & $113.67 \pm 3.09$ & $126.00 \pm 0.00$ & $102.33 \pm 2.62$ & $100.00 \pm 0.00$ & $81.22 \pm 2.08$ & $89.62 \pm 1.26$ \\
    \bottomrule
  \end{tabular*}
\end{table}

\subsection{The Trajectory-Linked Review Dataset}
\label{sec:trajectory-dataset}

The human PR corpus allows us to compare generated review comments with
human review comments, but it does not provide the development trajectories
needed to study trajectory-informed review. We therefore turn to DeepSWE.
Each task provides a user request, a base commit, an execution environment,
and a held-out reference implementation. These resources allow us to run
coding agents on the tasks, record their development trajectories, and
generate reference review comments using the procedure introduced in
Section~\ref{sec:reference-construction}.

We use all 113 DeepSWE tasks from 91 repositories: 34 Go, 34 Python,
35 TypeScript, 5 JavaScript, and 5 Rust tasks~\cite{datacurve2026deepswe}.
For each task, we retain one
implementation and development trajectory from each of three coding
configurations: GPT-5.5 with Codex CLI, GPT-5.6 Sol with Codex CLI,
and Gemini 3.8 Flash with Copilot CLI. This yields
\ImplementationCount{} implementations, with the coding task as the
sampling unit.

To generate reference review comments for these implementations, we use the
held-out DeepSWE reference patch in place of the merged-version patch
used for human PRs. Here, the reference patch provides a comparison
target rather than a recorded repair of the agent's implementation.
GPT-5.6 Sol at xhigh reasoning effort compares the original task
description, the agent's implementation patch, and the reference patch
to generate review comments about the agent's implementation.
This produces 355, 185, and 342 reference
review comments for the GPT-5.5, GPT-5.6, and Gemini collections, respectively,
for a total of \SyntheticLabelCount{} reference review comments.
In total, the dataset links \ImplementationCount{} implementations
produced by coding agents with their development trajectories and
\SyntheticLabelCount{} reference review comments, covering
113 tasks from 91 repositories.

\subsection{Study Design}
\label{sec:study-design}

\subsubsection{Experimental Setup}
\label{sec:review-execution}

We separate code generation from code review. As described in
Section~\ref{sec:trajectory-dataset}, three coding models independently
carry out the same set of tasks, producing three collections of code
changes and development trajectories.

We then review the code changes in each collection through the
\texttt{/review} modes of two review harnesses: Codex CLI and GitHub
Copilot CLI. Both harnesses use GPT-5.6 Sol as the reviewer model,
regardless of which model produced the code. Crossing the three coding
models with the two review harnesses yields six coding--review configurations.

Within each configuration, we compare review conditions that differ in
whether and how the development trajectory is provided to the reviewer.
In every condition, the reviewer can inspect the final patch and repository.
Diff-only review receives neither a development
trajectory nor a separate task description. Trajectory-informed review
receives the trajectory $\tau$ or information derived from it; the
original user request $u_0$ is available through this context
(Section~\ref{sec:motivation-trajectory}). For each task and coding
configuration, all review conditions use the same final patch and
reference review comments. This pairs the comparisons on the same code,
rather than comparing reviews of different implementations. The held-out
reference implementation, reference review comments, and hidden tests
are never provided to the reviewers.

RQ1 compares diff-only review with access to the full trajectory.
RQ2 compares alternative ways of representing and using the trajectory.
RQ3 examines their resource costs and recall of reference review comments
linked to critical task failures.

\subsubsection{Evaluation Metrics}
\label{sec:measures}
\label{sec:review-evaluation}

We compare the review comments produced by each reviewer with the fixed
reference review comments using semantic matching. GPT-5.6 Sol compares
comments concerning the same file. Two comments match when the code
change needed to address the concern in one would also resolve or
substantially address the concern in the other. One reviewer comment may match
several reference comments, and several reviewer comments may match the
same reference comment. Each reference comment is nevertheless counted
at most once.

Our primary metric is \emph{micro recall}, the proportion of reference
review comments matched by the reviewer's output across all tasks.
For a given review condition and coding--review configuration, let $G_i$
be the number of reference review comments for task $i$, and $C_i$ the
number matched by at least one reviewer comment. We compute:
\begin{equation}
  R_{\mathrm{micro}} =
  \frac{\sum_{i=1}^{113} C_i}{\sum_{i=1}^{113} G_i},
\end{equation}
This gives each reference comment equal weight, rather than averaging
per-task recalls. It measures coverage of the reference comments, not the
correctness of every comment produced by the reviewer.

We compute micro recall separately for each of the six coding--review
configurations. The overall mean recall is the arithmetic mean of the six
configuration-level micro recalls. An empty review contributes $C_i=0$.
A task with no reference comments has $G_i=C_i=0$ and contributes nothing
to the pooled counts, but its review remains included in review-output
and resource-use summaries.
% !TEX root = ../../../main.tex
\section{Results}
\label{sec:evaluation}

\subsection{RQ1: Does access to the full development trajectory improve code review over diff-only review?}
\label{sec:rq1}

We first test whether providing the full development trajectory $\tau$
helps a reviewer recover more reference review comments. Diff-only review
provides the final patch and repository, while full-trajectory review
additionally provides the complete development record without first
selecting or summarizing its contents. We compare these conditions within
each of the six coding--review configurations using the micro recall
defined in Section~\ref{sec:measures}.

Table~\ref{tab:full-trajectory-results} shows that full-trajectory review
increases mean reference recall from 23.03\% to 25.87\%, a gain of 2.83
percentage points. Recall improves in four of the six configurations.
The largest gain, 6.76 percentage points, occurs when GitHub Copilot CLI
reviews code produced by GPT-5.5. In contrast, recall decreases for code
produced by GPT-5.6 Sol under both review harnesses: by 1.08 percentage
points with Codex CLI and 0.54 percentage points with GitHub Copilot CLI.

% !TEX root = ../../../../main.tex
\begin{table}[t]
  \caption{Reference recall of diff-only and full-trajectory review.}
  \label{tab:full-trajectory-results}
  \compacttable
  \begin{tabularx}{\textwidth}{@{}XXrrr@{}}
    \toprule
    Coding model & Review harness & \shortstack{Diff only\\Recall (\%)} & \shortstack{Full trajectory\\Recall (\%)} & \shortstack{$\Delta R$\\(pp)} \\
    \midrule
    GPT-5.5 & Codex CLI & 27.89 & 31.55 & +3.66 \\
    GPT-5.6 Sol & Codex CLI & 16.22 & 15.14 & -1.08 \\
    Gemini 3.8 Flash & Codex CLI & 26.90 & 28.95 & +2.05 \\
    GPT-5.5 & GitHub Copilot CLI & 26.76 & 33.52 & +6.76 \\
    GPT-5.6 Sol & GitHub Copilot CLI & 16.76 & 16.22 & -0.54 \\
    Gemini 3.8 Flash & GitHub Copilot CLI & 23.68 & 29.82 & +6.14 \\
    \midrule
    \multicolumn{2}{@{}l}{Mean} & 23.03 & 25.87 & +2.83 \\
    \bottomrule
  \end{tabularx}
\end{table}

To understand how reviewers access the full trajectory, we examined the
text returned by their trajectory-reading tool calls. Only about 10\%
of the trajectory text was returned to the reviewer through these calls.
Thus, making the complete record available does not mean that the reviewer
accesses most of its contents. This observation motivates RQ2: can a
different representation make review-relevant information more accessible
and improve review effectiveness?

\answerbox{
Providing the full development trajectory improves mean reference recall
by 2.83 percentage points over diff-only review. However, recall improves
in only four of six configurations, so the benefit is not consistent
across coding models and review harnesses.
}

% !TEX root = ../../../main.tex
\subsection{RQ2: Which representations and review procedures best support trajectory-informed code review?}
\label{sec:rq2}
\label{sec:review-methods}

RQ1 shows that providing the full trajectory can improve review, but the
gains vary across configurations and reviewers access only part of the
available record. We next investigate whether different representations
and ways of using trajectory information can improve review effectiveness.
We compare four trajectory-informed methods with diff-only review,
first describing the methods, then comparing their recall, and finally
examining how they use development evidence.

\subsubsection{Review Methods}
\label{sec:context-challenges}
\label{sec:signal-families}
\label{sec:design-rationale}

Using the notation from Section~\ref{sec:motivation-trajectory}, the four
methods differ in which contents of $\tau$ they retain, how they organize
those contents, and when the reviewer uses them.

\paragraph{Messages.}
Messages retains all user and agent text in chronological, role-labeled
order, including the user request $u_0$, agent messages $m_i$, and the
completion message $m_{\mathrm{final}}$. It omits structured tool calls
$c_i$ and observations $o_i$. The reviewer checks the retained account against
the patch and repository.

\paragraph{Self-selected signals.}
The reviewer receives the full normalized trajectory $\tau$. Before
inspecting code, it extracts requirements from $u_0$ and selects evidence
from the messages $m_i$, tool calls $c_i$, and observations $o_i$ that it
considers useful for review.
There is no fixed output schema or question quota; the reviewer decides
what to retain and how to organize it.

\paragraph{Evidence-linked map.}
\label{sec:evidence-map}
The coordinator receives the full trajectory $\tau$ and, before inspecting
code, extracts requirements from $u_0$ and creates a single evidence-linked map.
Facts and questions cite their source events in
$\tau$, such as messages $m_i$, calls $c_i$, or observations $o_i$, and
distinguish observations from interpretations. Each prioritized question connects
evidence to a code location, a plausible counterexample, and a focused check.

The map is fixed before investigation. The reviewer checks its leads
against the patch, callers, consumers, and tests, with a limit of ten
focused probes; broad suites and repeated checks are discouraged.
Delegated workers return code evidence without changing the map, and an
independent code-driven pass supplements the map's leads.

\paragraph{Code-first adversarial scrutiny.}
The reviewer first inspects the patch and repository independently of
the development trajectory. It then consults $\tau$ to challenge its initial
assessment: implementation assumptions and safety claims in $m_i$, together
with validation results in $o_i$, become evidence for questions to test.
The review must
report concrete code problems rather than complaints about plans or
workflow compliance. Unlike Messages, this method does not reduce the
available record; it changes when and how the reviewer uses it.

\subsubsection{Comparison Results}

We first compare the methods by their mean reference recall across the six
coding--review configurations (Table~\ref{tab:review-means}).
The evidence-linked map method performs best overall, reaching 30.35\% compared
with 23.03\% for diff-only, an improvement of 7.32 percentage points.
Messages is second at 29.03\%. All four methods improve over diff-only,
with gains ranging from 4.05 to 7.32 percentage points.
Their mean recalls also exceed the 25.87\% obtained by full-trajectory
review in RQ1.

% !TEX root = ../../../../main.tex
\begin{table}[t]
  \caption{Mean reference recall and gains over diff-only.}
  \label{tab:review-means}\compacttable
  \begin{tabular*}{\textwidth}{@{\extracolsep{\fill}}lrr@{}}
    \toprule
    Condition & Recall (\%) & $\Delta R$ (pp) \\
    \midrule
    Diff only & 23.03 & --- \\
    Messages & 29.03 & +6.00 \\
    Self-selected signals & 27.83 & +4.80 \\
    Evidence-linked map & 30.35 & +7.32 \\
    Adversarial scrutiny & 27.08 & +4.05 \\
    \bottomrule
  \end{tabular*}
\end{table}

We next examine whether the same ranking holds within each configuration
(Table~\ref{tab:review-results}). The evidence-linked map method has the highest recall in four
configurations and ties for highest in a fifth: when Codex CLI reviews
GPT-5.6 Sol code, both the evidence-linked map method and adversarial scrutiny reach 18.92\%.
The remaining configuration reverses the ranking of the two strongest
methods overall. When GitHub Copilot CLI reviews GPT-5.6 Sol code,
Messages reaches 21.62\%, while the evidence-linked map method reaches 18.38\%.
Thus, the method's advantage is observed across most configurations,
but a simpler message view performs better in one.
Together, these results favor the evidence-linked map method on average while
showing that Messages remains competitive.

% !TEX root = ../../../../main.tex
\begin{table*}[tp]
  \caption{Reference recall and differences from diff-only.}
  \label{tab:review-results}
  \compacttable[0.90]
  \begin{tabular*}{\textwidth}{@{\extracolsep{\fill}}lllrrrr@{}}
    \toprule
    \shortstack[l]{Coding\\model} & \shortstack[l]{Review\\harness} & Condition & \shortstack{Recall\\(\%)} & \shortstack{Report\\candidates} & \shortstack{Covered\\references} & \shortstack{$\Delta R$\\(pp)} \\
    \midrule
    % configuration A
    \multirow{5}{*}{\shortstack[l]{GPT-5.5}} & \multirow{5}{*}{\shortstack[l]{Codex CLI}} & Evidence-linked map & \textbf{36.34} & 394 & 129 & +8.45 \\
     &  & Messages & 36.06 & 370 & 128 & +8.17 \\
     &  & Self-selected signals & 35.49 & 364 & 126 & +7.61 \\
     &  & Adversarial scrutiny & 33.24 & 389 & 118 & +5.35 \\
     &  & Diff only & 27.89 & 376 & 99 & --- \\
    \midrule
    % configuration B
    \multirow{5}{*}{\shortstack[l]{GPT-5.6 Sol}} & \multirow{5}{*}{\shortstack[l]{Codex CLI}} & Adversarial scrutiny & \textbf{18.92} & 228 & 35 & +2.70 \\
     &  & Evidence-linked map & \textbf{18.92} & 219 & 35 & +2.70 \\
     &  & Self-selected signals & 17.84 & 189 & 33 & +1.62 \\
     &  & Messages & 16.76 & 204 & 31 & +0.54 \\
     &  & Diff only & 16.22 & 244 & 30 & --- \\
    \midrule
    % configuration C
    \multirow{5}{*}{\shortstack[l]{Gemini 3.8 Flash}} & \multirow{5}{*}{\shortstack[l]{Codex CLI}} & Evidence-linked map & \textbf{36.26} & 413 & 124 & +9.36 \\
     &  & Messages & 30.41 & 362 & 104 & +3.51 \\
     &  & Self-selected signals & 29.24 & 361 & 100 & +2.34 \\
     &  & Adversarial scrutiny & 28.36 & 403 & 97 & +1.46 \\
     &  & Diff only & 26.90 & 378 & 92 & --- \\
    \midrule
    % configuration D
    \multirow{5}{*}{\shortstack[l]{GPT-5.5}} & \multirow{5}{*}{\shortstack[l]{GitHub\\Copilot CLI}} & Evidence-linked map & \textbf{38.31} & 392 & 136 & +11.55 \\
     &  & Messages & 37.46 & 361 & 133 & +10.70 \\
     &  & Self-selected signals & 34.08 & 344 & 121 & +7.32 \\
     &  & Adversarial scrutiny & 33.52 & 422 & 119 & +6.76 \\
     &  & Diff only & 26.76 & 343 & 95 & --- \\
    \midrule
    % configuration E
    \multirow{5}{*}{\shortstack[l]{GPT-5.6 Sol}} & \multirow{5}{*}{\shortstack[l]{GitHub\\Copilot CLI}} & Messages & \textbf{21.62} & 217 & 40 & +4.86 \\
     &  & Adversarial scrutiny & 18.92 & 278 & 35 & +2.16 \\
     &  & Evidence-linked map & 18.38 & 206 & 34 & +1.62 \\
     &  & Self-selected signals & 17.30 & 199 & 32 & +0.54 \\
     &  & Diff only & 16.76 & 251 & 31 & --- \\
    \midrule
    % configuration F
    \multirow{5}{*}{\shortstack[l]{Gemini 3.8 Flash}} & \multirow{5}{*}{\shortstack[l]{GitHub\\Copilot CLI}} & Evidence-linked map & \textbf{33.92} & 370 & 116 & +10.23 \\
     &  & Self-selected signals & 33.04 & 350 & 113 & +9.36 \\
     &  & Messages & 31.87 & 376 & 109 & +8.19 \\
     &  & Adversarial scrutiny & 29.53 & 425 & 101 & +5.85 \\
     &  & Diff only & 23.68 & 353 & 81 & --- \\
    \bottomrule
  \end{tabular*}
\end{table*}

\paragraph{Complementary reference coverage.}
\label{sec:output-overlap}
Higher recall does not mean that one method recovers every reference
comment found by another. Table~\ref{tab:review-overlap} compares the
comments matched by the evidence-linked map method and diff-only review.
For GPT-5.5 code under Codex CLI, both methods match 71 comments;
another 58 are matched only by the evidence-linked map method, and 28
only by diff-only. Diff-only matches comments missed by the evidence-linked
map method in every configuration.

% !TEX root = ../../../../main.tex
\begin{table}[t]
  \caption{Map and diff-only reference overlap.}
  \label{tab:review-overlap}\compacttable
  \begin{tabular*}{\textwidth}{@{\extracolsep{\fill}}llrrrr@{}}
    \toprule
    Coding model & Reviewer & Map only & Shared & Diff only & Jaccard \\
    \midrule
    GPT-5.5 & Codex CLI /review & 58 & 71 & 28 & 0.452 \\
    GPT-5.6 Sol & Codex CLI /review & 18 & 17 & 13 & 0.354 \\
    Gemini 3.8 Flash & Codex CLI /review & 60 & 64 & 28 & 0.421 \\
    GPT-5.5 & Copilot CLI /review & 67 & 69 & 26 & 0.426 \\
    GPT-5.6 Sol & Copilot CLI /review & 19 & 15 & 16 & 0.300 \\
    Gemini 3.8 Flash & Copilot CLI /review & 64 & 52 & 29 & 0.359 \\
    \bottomrule
  \end{tabular*}
\end{table}

To check whether the observed differences persist across repeated runs,
we run all six methods, including diff-only and full-trajectory review,
three times independently on GPT-5.5 code reviewed through GitHub Copilot CLI.
Table~\ref{tab:repeated-run-recall} reports each round's recall and the
three-round mean. The overall pattern is stable across runs: all trajectory-informed methods consistently outperform diff-only, while the variation across repetitions remains small. In particular, Messages and the evidence-linked map achieve the highest mean recall at 36.71\%, compared with 27.14\% for diff-only.

\begin{table}[t]
  \caption{Reference recall (\%) across three independent runs on GPT-5.5
  implementations with GitHub Copilot CLI \texttt{/review}.}
  \label{tab:repeated-run-recall}
  \compacttable
  \begin{tabular*}{\textwidth}{@{\extracolsep{\fill}}lrrrr@{}}
    \toprule
    Method & Round 0 & Round 1 & Round 2 & Mean $\pm$ SD \\
    \midrule
    Diff only & 26.76 & 26.20 & 28.45 & $27.14 \pm 0.96$ \\
    Full trajectory & 33.52 & 34.37 & 36.34 & $34.74 \pm 1.18$ \\
    Messages & 37.46 & 36.34 & 36.34 & $36.71 \pm 0.53$ \\
    Self-selected signals & 34.08 & 34.08 & 34.08 & $34.08 \pm 0.00$ \\
    Evidence-linked map & 38.31 & 36.06 & 35.77 & $36.71 \pm 1.13$ \\
    Adversarial scrutiny & 33.52 & 36.06 & 33.52 & $34.37 \pm 1.20$ \\
    \bottomrule
  \end{tabular*}
\end{table}

\subsubsection{Information Retained and Used for Review}
\label{sec:information-extraction}

We inspect the generated evidence-linked maps to understand whether their
questions draw on the development process rather than only restating the
user request $u_0$. We automatically analyze the structure and source identifiers
of \InfoMapCount{} maps, covering \ImplementationCount{} implementations
reviewed through two harnesses. The maps contain \InfoQuestionCount{}
questions, with a median of \InfoQuestionMedian{} per map.
Of the questions, \InfoBeyondRequestCount{} (\InfoBeyondRequestPercent{}\%)
cite events beyond $u_0$, and \InfoToolBearingCount{}
(\InfoToolBearingPercent{}\%) cite events containing tool calls $c_i$
or observations $o_i$.
The questions therefore cite development evidence beyond the task
description.
We examine the information recorded in the generated maps across seven
categories, summarized in Table~\ref{tab:trajectory-information}.

% !TEX root = ../../../../main.tex
% Generated by scripts/update-trajectory-information.mjs; categories for analyzing map contents.
\begin{table*}[t]
  \caption{Seven categories used to examine information in generated evidence-linked maps.}
  \label{tab:trajectory-information}
  \compacttable
  \begin{tabularx}{\textwidth}{@{}p{0.38\textwidth}X@{}}
    \toprule
    Information category & Information recorded in maps \\
    \midrule
    Implementation strategy and evolution & The chosen implementation approach and how it changed. \\
    Requirements and implementation decisions & Requested behavior and decisions about how to implement it. \\
    Explicit development assumptions & Stated assumptions about inputs, behavior, or compatibility. \\
    Failures and revisions & Failed attempts or checks and the subsequent changes. \\
    Validation evidence and limits & Checks performed, their outcomes, and stated limits. \\
    Changed artifacts and consumers & Modified files or interfaces and their affected callers or consumers. \\
    Unresolved or conflicting information & Open questions, conflicting observations, or unfinished work. \\
    \bottomrule
  \end{tabularx}
\end{table*}

\paragraph{Example: an unspecified option becomes disabled.}
FastAPI\footnote{\url{https://deepswe.datacurve.ai/data/v1.1/tasks/fastapi-implicit-head-options}}
groups API endpoints into routers, which can be nested inside other
routers. The task adds settings for automatic responses to HTTP HEAD and
OPTIONS requests. We focus on an OPTIONS bug: the code turns an unset
option into an explicit \texttt{False}, preventing an outer setting from
enabling the response.

The option has three states: enabled, disabled, and unspecified.
An unspecified inner option should allow an explicit outer setting to
take effect. In particular,
\emph{unspecified} must not be treated as an explicit \emph{disabled}.

The development history suggests where to investigate this distinction.
The coding agent had already encountered a test failure involving which
setting takes precedence. After a repair, all 11 tests in its targeted
suite passed:

\tracebox{Source trajectory: failing check and subsequent pass}{\begin{lstlisting}
[test output before repair (excerpt)]
>       assert client.options("/p2/r2-inc-win").status_code == 200
E       AssertionError: assert 405 == 200

[test output after repair (excerpt)]
11 passed
\end{lstlisting}
}

The evidence-linked map uses that history to ask whether the repaired
logic also works when routers are nested or reused. It identifies
\texttt{include\_router}, which copies one router's routes into another.
The proposed check includes the same child through two parents with
different settings and tests both HEAD and OPTIONS:

\par\smallskip\noindent
\begingroup\setlength{\fboxsep}{1pt}%
\fbox{\begin{minipage}{\dimexpr\linewidth-2\fboxsep-2\fboxrule\relax}\small
\textbf{Saved evidence-linked map question (excerpt)}\par\smallskip
\begin{tabularx}{\linewidth}{@{}lX@{}}
  \textbf{Evidence} & inclusion copies routes and resolves route/include/router/self, with a later correction to precedence logic. \\
  \textbf{Question} & Does repeated/nested inclusion preserve nearest non-omitted precedence after routes have already been materialized? \\
  \textbf{Surface} & \texttt{include\_router}. \\
  \textbf{Counterexample} & same child router included through two parents with opposite include-level settings. \\
  \textbf{Probe} & nested two-parent inclusion matrix for both HEAD and OPTIONS. \\
  \textbf{Source example} & \texttt{97fa8712-3cfb-4dbd-8169-b0b4c0bde2ea} \\
\end{tabularx}
\end{minipage}}
\endgroup\par\smallskip

The reviewer tests the same precedence rule in a simpler nested
configuration: a child router is included in an intermediate router, with
OPTIONS left unspecified at both levels and enabled only at the outer
inclusion. The outer setting should therefore enable the OPTIONS response.
Instead, the application returns HTTP 405
(\emph{Method Not Allowed}). The probe also shows that the copied route
already stores \texttt{False}:

\tracebox{Recorded review probe result}{\begin{lstlisting}
PROBE2 nested omitted then outer include override
405 {"detail":"Method Not Allowed"} materialized flags [False]
\end{lstlisting}
}

The source patch explains the failure in two steps.
\emph{At the first inclusion}, the child route is copied into the
intermediate router. Because no setting is supplied, \texttt{default=False}
converts the unspecified option into \texttt{False}.
\emph{At the next inclusion}, the code gives this stored route-level
value priority over the outer \texttt{True}. It can no longer distinguish
an explicit decision to disable OPTIONS from an option that was never set:

\patchbox{Source patch: \texttt{fastapi/routing.py} inclusion precedence}{\begin{lstlisting}
+auto_options=resolve_nearest(
+    route.auto_options,
+    auto_options,
+    router.auto_options,
+    default=False,
+),
\end{lstlisting}
}

This example illustrates how an evidence-linked map turns a recorded
development failure and its subsequent repair into a concrete review
question. The trajectory guides what to investigate, while the review
probe and final code establish whether the suspected problem remains.

% \subsubsection{Outcome-stratified comparison}
% \label{sec:outcome-analysis}

% Table~\ref{tab:outcome-analysis} averages the six within-group Micro recalls.
% Its last column is failed minus passed. The map's mean recall in the failed
% group is 31.98\%, an
% 8.57-point gain
% over diff-only. The corresponding gain in the passed group is
% 6.05 points. This larger improvement on failed implementations suggests that evidence-linked review is particularly useful when the coding agent introduces functional deviations, making such defects more likely to be surfaced during review.

\answerbox{
The evidence-linked map achieves the highest overall mean recall, outperforming diff-only by 7.32 percentage points. Repeated runs further show that trajectory-informed methods consistently outperform diff-only, with Messages and the evidence-linked map tying for the highest mean recall.
}

% !TEX root = ../../../main.tex
\subsection{RQ3: What are the impacts of trajectory-informed review on cost and critical failure-linked coverage?}
\label{sec:rq3}
\label{sec:coverage-resource}

RQ2 shows that trajectory-informed methods recover more reference review
comments on average. We now examine whether these gains extend to comments
that explain task failures, and what additional resources the methods
require. We compare the same five review conditions across the six
coding--review configurations.

\subsubsection{Failure-Linked Reference Recall}

We distinguish reference review comments that explain why an implementation
fails its task from those that raise other concerns. GPT-6 Astra examines
each reference review comment together with the associated DeepSWE task
and the coding agent's execution outcome. A comment is labeled
\emph{critical failure-linked} when the reported problem directly explains
an official DeepSWE verifier failure. The \emph{other} group includes
comments unrelated to that failure and comments on implementations that
pass the verifier.

The analysis includes 51 critical failure-linked and 829 other
reference review comments. The 2 references associated with a verifier
timeout are excluded from this comparison. Within each group, we compute
micro recall separately for each configuration, following
Section~\ref{sec:measures}, then take the arithmetic mean of the six values.

% !TEX root = ../../../../main.tex
\begin{table}[t]
  \caption{Reference recall by functional relevance (\%; six-configuration mean).}
  \label{tab:review-severity}\compacttable
  \begin{tabular*}{\textwidth}{@{\extracolsep{\fill}}lrr@{}}
    \toprule
    Condition & Critical failure-linked & Other \\
    \midrule
    Diff only & 22.90 & 22.79 \\
    Messages & 30.81 & 28.59 \\
    Self-selected signals & 31.65 & 27.40 \\
    Evidence-linked map & 40.40 & 29.73 \\
    Adversarial scrutiny & 32.49 & 26.47 \\
    \bottomrule
  \end{tabular*}
\end{table}

Table~\ref{tab:review-severity} shows that the evidence-linked map method
has the highest mean recall in both groups. For critical failure-linked
comments, recall increases from 22.90\% with diff-only to
40.40\%, a gain of 17.51 percentage points. For other
comments, it increases from 22.79\% to 29.73\%, a gain of
6.94 percentage points. The larger gain on failure-linked comments
shows that the improvement is not confined to concerns unrelated to task
failure. The other three trajectory-informed methods also exceed diff-only
on this group.

\subsubsection{Resource Use and Cost}

For each method and configuration, we report median input tokens, output
tokens, elapsed review time, and estimated API cost over 113 selected
review sessions. The pooled rows combine all five conditions, yielding
565 sessions per configuration. These measurements cover the entire review
session, including preparing trajectory-derived information and executing
delegated review tasks, but exclude the preceding coding and
reference-generation stages.

Input counts include uncached input, cache reads, and cache writes;
output counts include reasoning output. Time spans recorded start to
finish, including tool execution. Parenthesized values in
Table~\ref{tab:review-token-usage} are cache-read token medians, computed
separately from total-input medians. Cache reads are already included in
total input and should not be added to it.

Both review harnesses use GPT-5.6 Sol. We estimate API-equivalent token
charges using its standard prices archived on
2026-09-26.\footnote{OpenAI API pricing:
\url{https://developers.openai.com/api/docs/pricing}.}
We price ordinary input, cache reads, cache writes, and output per
request, sum within each review, then take the median.

% !TEX root = ../../../../main.tex
\begin{table}[t]
  \caption{Median input/output tokens (thousands), review time (minutes), and estimated API cost (USD). Parenthesized values report the median number of input tokens read from cache; these tokens are included in total input.}
  \label{tab:review-token-usage}\compacttable\scriptsize\setlength{\tabcolsep}{2pt}
  \begin{tabular*}{\textwidth}{@{\extracolsep{\fill}}lrrrrrrrrrrrr@{}}
    \toprule
    \multicolumn{13}{@{}l}{\textit{Codex CLI /review}} \\
    & \multicolumn{4}{c}{GPT-5.5} & \multicolumn{4}{c}{GPT-5.6 Sol} & \multicolumn{4}{c}{Gemini 3.8 Flash} \\
    Condition & Input (cached) & Output & Time & Cost & Input (cached) & Output & Time & Cost & Input (cached) & Output & Time & Cost \\
    \midrule
    Diff only & 343.9 (251.6) & 4.8 & 1.85 & 0.56 & 362.0 (313.7) & 4.6 & 2.18 & 0.41 & 408.2 (336.6) & 4.9 & 2.08 & 0.51 \\
    Messages & 271.3 (187.4) & 6.0 & 1.88 & 0.54 & 350.5 (284.7) & 5.7 & 2.43 & 0.45 & 441.6 (366.7) & 6.1 & 2.22 & 0.60 \\
    Self-selected signals & 454.3 (314.6) & 7.0 & 2.23 & 0.79 & 501.3 (454.3) & 7.0 & 2.67 & 0.60 & 579.1 (481.3) & 7.3 & 2.58 & 0.77 \\
    Evidence-linked map & 964.9 (758.0) & 11.5 & 3.22 & 1.38 & 1140.4 (1008.8) & 11.4 & 3.75 & 1.13 & 1545.4 (1309.1) & 14.3 & 4.08 & 1.66 \\
    Adversarial scrutiny & 654.6 (501.2) & 7.1 & 2.57 & 0.96 & 734.0 (562.2) & 7.5 & 2.72 & 1.06 & 799.3 (683.8) & 7.7 & 2.87 & 0.94 \\
    \midrule
    All five conditions & 498.4 (361.0) & 6.8 & 2.30 & 0.81 & 580.3 (497.1) & 6.8 & 2.72 & 0.71 & 645.8 (548.2) & 7.0 & 2.72 & 0.81 \\
    \midrule
    \multicolumn{13}{@{}l}{\textit{Copilot CLI /review}} \\
    & \multicolumn{4}{c}{GPT-5.5} & \multicolumn{4}{c}{GPT-5.6 Sol} & \multicolumn{4}{c}{Gemini 3.8 Flash} \\
    Condition & Input (cached) & Output & Time & Cost & Input (cached) & Output & Time & Cost & Input (cached) & Output & Time & Cost \\
    \midrule
    Diff only & 176.3 (124.7) & 4.9 & 2.05 & 0.38 & 207.6 (144.9) & 5.0 & 1.97 & 0.46 & 210.9 (172.5) & 5.1 & 2.42 & 0.36 \\
    Messages & 205.0 (131.1) & 5.8 & 2.20 & 0.46 & 277.8 (176.1) & 6.7 & 2.27 & 0.60 & 248.4 (200.1) & 6.3 & 2.90 & 0.46 \\
    Self-selected signals & 340.1 (224.6) & 7.4 & 2.60 & 0.66 & 435.2 (256.5) & 8.2 & 2.72 & 0.91 & 373.8 (299.5) & 7.4 & 3.38 & 0.60 \\
    Evidence-linked map & 1090.5 (813.6) & 15.1 & 5.58 & 1.70 & 1273.6 (840.7) & 15.5 & 5.25 & 2.06 & 1195.9 (1067.3) & 17.5 & 7.12 & 1.54 \\
    Adversarial scrutiny & 488.1 (413.8) & 7.8 & 3.40 & 0.71 & 573.7 (473.1) & 8.1 & 3.70 & 0.75 & 525.2 (445.2) & 8.3 & 4.05 & 0.73 \\
    \midrule
    All five conditions & 343.5 (255.5) & 7.2 & 2.87 & 0.64 & 435.2 (311.3) & 7.8 & 3.13 & 0.75 & 398.1 (316.1) & 7.4 & 3.70 & 0.61 \\
    \bottomrule
  \end{tabular*}
\end{table}

The evidence-linked map method has the largest input and output token
medians in all six configurations: 2.81--6.18 times
diff-only's input and 2.39--3.45 times its output.
Its median review time ranges from 3.22--7.12 minutes,
compared with 1.85--2.42 minutes for diff-only.
Estimated median API cost ranges from USD 1.13--2.06 per review,
compared with USD 0.36--0.56 for diff-only.

Messages offers a lower-cost alternative to the evidence-linked map method.
It uses fewer input and output tokens in every configuration, with median
review times of 1.88--2.90 minutes and estimated median costs
of USD 0.45--0.60. Its mean recall on critical failure-linked
comments is 30.81\%, above diff-only's
22.90\% but below the evidence-linked map method's
40.40\%. Thus, the evidence-linked map method offers higher
recall when the additional time and token use are acceptable, while Messages
reduces this overhead at the cost of lower recall.

\answerbox{
The evidence-linked map method achieves the highest mean recall on both
critical failure-linked and other reference review comments, improving
over diff-only by 17.51 and 6.94 percentage points, respectively.
It also requires the most tokens and the longest review time.
Messages offers a lighter-weight alternative: it improves failure-linked
recall over diff-only while using fewer resources than the evidence-linked
map method.
}

% !TEX root = ../../../main.tex
\section{Related Work}
\label{sec:related-work}

\subsection{Learning-Based Code Review}

Automated code review has evolved from task-specific learned models toward
large language model (LLM) systems with increasingly rich sources of
context. Early work focused primarily on improving the review model itself.
CodeReviewer introduced large-scale pre-training tailored to code review,
covering code-change quality estimation, review-comment generation, and
code refinement~\cite{li2022codereviewer}. Subsequent work adapted
general-purpose LLMs with parameter-efficient or task-specific
fine-tuning. LLaMA-Reviewer uses parameter-efficient tuning of
LLaMA models~\cite{lu2023llamareviewer}; Pornprasit and
Tantithamthavorn compare fine-tuning with zero-shot, few-shot, and
persona-based prompting~\cite{pornprasit2024finetuning}; and Haider
et al.\ combine prompt augmentation with parameter-efficient
fine-tuning~\cite{haider2024prompting}. CARLLM jointly targets the accuracy
and comprehensibility of generated reviews~\cite{yu2024carllm}, while
MelcotCR uses long chain-of-thought fine-tuning and structured review
dimensions to strengthen multi-dimensional review
reasoning~\cite{yu2025melcotcr}. These approaches improve the reviewing
capability encoded in the model, but the information available at inference
time remains largely determined by the code change and the context supplied
to the reviewer.

Rasheed et al.\ provide an early multi-agent example, decomposing GPT-4-based
review across agents for general review, bug detection, code smells, and
optimization~\cite{rasheed2024aipoweredreview}. Their reported evidence is
qualitative and preliminary, with systematic comparisons and developer-facing
evaluation left to future work. This illustrates a broader distinction:
improving or orchestrating the review model does not determine which process
evidence the reviewer receives, and review quality may require evaluation
beyond reference-finding recall alone.

\subsection{Context-Augmented Code Review}

A complementary line of work enriches review with additional context.
AUGER retrieves relevant code changes and review comments to support
comment generation~\cite{li2022auger}, and RAG-Reviewer conditions
generation on similar historical code--review pairs~\cite{hong2025ragreviewer}.
More recent systems broaden the available evidence. LAURA combines
retrieved review exemplars with pull-request metadata and code
context~\cite{zhang2025laura}, while context-aware retrieval-augmented
review grounds comments in repository evidence~\cite{icoz2026contextaware}.
Sun et al.\ systematically study neighboring code, language-server-derived
semantic context, and similar co-change context, finding that contextual
augmentation can improve the coverage of an issue-list-based review
pipeline~\cite{sun2026gocodereview}. Haider et al.\ likewise augment
patches with semantic information such as call graphs and code
summaries~\cite{haider2024prompting}. At the evaluation level, AACR-Bench
preserves repository-level context and expert-verified findings, including
issues that require reasoning beyond changed lines~\cite{zhang2026aacrbench}.
Together, these studies show that a final diff may be insufficient and
that additional context can expose otherwise difficult-to-recover concerns.
Their context, however, is primarily reconstructed from the repository,
metadata, or historical examples at review time.

\subsection{Agentic Code Review}

More recently, code review has adopted agentic workflows in which reviewers
actively gather and exchange evidence. CodeAgent decomposes review among
communicating specialized agents coordinated by a supervisory
agent~\cite{tang2024codeagent}. Other multi-agent frameworks combine
review, debugging, and security-analysis agents with FAISS-based retrieval
memory~\cite{sharanarthi2025multiagent}. OpenCodeReview equips agents with
repository-exploration tools, file-level subagents, and an independent
reflection stage, allowing them to recover cross-file evidence rather than
relying on a fixed retrieved context~\cite{li2026opencodereview}. This
direction reflects the growing use of general coding agents for review.
Kamal{\i} et al.\ broaden this direction into an end-to-end vision spanning
PR creation, augmentation, reviewer selection, AI-assisted review, and
retrospective learning, with specialized agents and explicit human quality
gates~\cite{kamali2026agenticvision}. Their lifecycle framing proposes where
runtime traces and retained review evidence should flow; our evidence-linked
map operationalizes one narrower mechanism for preserving development-process
evidence as source-linked review questions.
However, an empirical study of 3,109 pull requests with review-state data
reports substantial noise in agent-generated feedback, indicating that
greater autonomy alone does not guarantee useful
findings~\cite{chowdhury2026industry}.

Recent empirical studies examine agentic review behavior and its practical
outcomes. Charoenwet et al.\ analyze terminal-based review trajectories,
showing that reviewers differ substantially in how they allocate exploration,
validation, and cost, and arguing for trajectory-aware, cost-sensitive
evaluation~\cite{charoenwet2026terminalreview}. Lin et al.\ study 31,073
CodeRabbit review--feedback pairs and find that developers reject 56.3\% of
the comments, with rejections including false positives, redundant or
out-of-scope feedback, and misalignment with developer intent or project
practice~\cite{lin2026coderabbit}. This rejection rate is not equivalent to
a false-positive rate, but it provides in-the-wild evidence that comment
volume does not imply practical usefulness. At the project level, Zhong
et al.\ analyze 1.02 million reviewed pull requests from 207 projects across
human-centric, LLM-assisted, and agentic review eras. Agent involvement is
associated with faster decisions in some adoption patterns, but not with
better review quality~\cite{zhong2026agenticreview}. These studies complement
our controlled reference-coverage and cost comparisons with behavioral and
developer-response evidence from deployed review workflows.

Related work also studies agents as code authors. Njoku et al.\ find that
agent-authored and human-authored pull requests differ in integration
outcomes, task sensitivity, and review communication~\cite{njoku2026codeauthors}.
These socio-technical differences reinforce that reviewing agent-authored
changes involves more than evaluating the final code artifact.

\subsection{Trajectory-Based Agent Evaluation}

Development trajectories have also become an evaluation target for coding
agents. Podivilov et al.'s AgentLens assesses complete interactive
trajectories using formal checks together with LLM-written trajectory
reviews and comparisons, enabling diagnoses that final-state scoring alone
cannot provide~\cite{podivilov2026agentlens}. A distinct system with the same
name by Sahoo et al.\ studies the ``Lucky Pass'' problem: among passing
trajectories in its evaluation subset, 10.7\% exhibit deficient processes
such as blind retries, missing verification, or disordered work despite
eventual task success~\cite{sahoo2026agentlens}. These works use trajectories
to evaluate the coding agent and the process that produced a solution. Our
work instead supplies that process evidence to an independent review agent
and measures whether it helps recover defects in the resulting code. The
distinction is important for our functional-relevance analysis: verifier
success or failure characterizes task outcomes, whereas our critical failure-linked
group identifies concrete residual defects that explain observed failures.

Abazar et al.\ identify agent-specific edit, rewrite, and repetition
signatures in pull-request patches from five coding
agents~\cite{abazar2026behavioral}. Their measurements characterize patch-level
behavior rather than chronological tool-use trajectories, but they support
the premise that development artifacts retain behavioral signals about how
agents produce changes. We study whether richer, ordered process evidence
can be converted into actionable leads for a separate reviewer.
% !TEX root = ../../../main.tex
\section{Threats to Validity}
\label{sec:threats}

\subsection{Construct Validity}
Reference recovery covers one aspect of code
review. Synthetic labels may be
incorrect, duplicated, or incomplete. A reference patch can embody one
valid design. Critical failure-linked classification also depends on the verifier's test
coverage. However, in this study, we rely on the correctness of DeepSWE reference patches, given that DeepSWE is a widely used benchmark for coding agents. In addition, our Human-Reference Check provides empirical validation for the effectiveness of our reference-finding construction procedure. Together, these factors give us confidence in the overall reliability of the constructed dataset.

\subsection{Internal Validity}
In the evaluation part, we do not report precision for reconstructed review comments. Some comments produced by the review agents may therefore be incorrect, unsupported, or unrelated to the reference findings. Code review is inherently open-ended: a comment not covered by the reference set may still identify a valid concern, and the validity of a finding can sometimes depend on design assumptions or interpretation. Since our reference findings are constructed around defects represented by the reference patch rather than an exhaustive enumeration of all valid review comments, treating unmatched comments as false positives would provide a misleading estimate of precision. We therefore focus our evaluation on reference recovery, measuring how effectively each review condition recovers known findings represented by the reference patch. Consequently, our results characterize differences in reference-finding coverage rather than the overall correctness or noise level of all generated review comments.

\subsection{External Validity}
Our evaluation covers three coding configurations and two review runtimes, but results may differ for other model families, agent architectures, or reviewer harnesses. In practice, however, developers tend to rely on a relatively small set of mature coding agents, model families, and review runtimes. We therefore believe that the combinations evaluated in this study cover a broad range of common practical settings.

% !TEX root = ../../../main.tex
\section{Ethical Considerations}
\label{sec:ethics}

\subsection{Data Privacy}

The DeepSWE experiments were conducted in virtual environments and did not
involve the collection or use of users' personal information. Our trajectory
analysis focuses on agent behavior and code-review outcomes. No evaluation
or profiling of individual developers was performed.

\subsection{Security and Enterprise Deployment}

The benchmark setting does not establish that unrestricted access to
trajectories is appropriate in enterprise environments. Production
deployments should preserve repository permissions and enforce
least-privilege access controls for both trajectories and derived
artifacts. Repository-, path-, and field-level filtering should exclude
restricted content before it reaches a reviewer or is exported.

For deployments involving sensitive repositories, secret scanning and
sanitization should precede analysis, dataset construction, or sharing,
and should cover both trajectories and derived artifacts. Where personal
identifiers are present, identity removal and appropriate anonymization
or pseudonymization should be applied; any identity mapping should be
stored separately under restricted access.

% !TEX root = ../../../main.tex
\section{Conclusion}
\label{sec:conclusion}

We study whether the development trajectory that produces a patch can
provide useful evidence for code review beyond the final diff. To enable
this investigation, we construct a trajectory-linked dataset that aligns
agent implementations, development trajectories, and reference findings,
and evaluate trajectory-informed review across multiple coding
configurations and reviewer harnesses. Providing the raw trajectory
increases mean reference recall across the six configurations from 23.03\%
to 25.87\%, a
gain of 2.83 percentage points, although the effect varies across
configurations. This variation shows that making development history
available is useful but not sufficient: reviewers also need effective ways
to identify and organize the relevant evidence.

Across the evaluated representations and review procedures, all four
trajectory-informed methods improve mean reference recall over diff-only
review by 4.05--7.32 percentage points. The evidence-linked map, which
organizes development evidence into source-linked review questions, leads
or ties in five of six configurations and achieves the largest mean gain.
It also exceeds diff-only review by 17.51 percentage points on critical failure-linked
references and by 6.94 percentage points on other references. These
improvements require additional tokens and review time, with Messages
providing a lighter-weight alternative when resource use is a primary
concern. Overall, our results show that development trajectories are a
valuable review artifact, but their value depends on transforming process
evidence into concrete questions whose validity is independently checked
against the final code. Evidence-linked map that retaining and structuring this evidence can help
reviewers recover implementation assumptions, incomplete validation, and
fragile behaviors that are difficult to reconstruct from the final patch
alone.

% !TEX root = ../../../main.tex
\section*{Data Availability}
\label{sec:data-validity}
All data for this research are available in the artifact on Hugging Face:
\par{\centering\url{https://huggingface.co/datasets/ZijiantheJack/ProcessAwareCodeReview}\par}

\label{paper:main-end}

\bibliographystyle{ACM-Reference-Format}
\bibliography{references,references-related}

\end{document}